\documentclass[conference,comsoc]{IEEEtran}  % journal or conference
\IEEEoverridecommandlockouts
\usepackage{cite} % by default: join 3 or more citations like [1], [2], [3] to [1-3].
\usepackage{amsmath,amssymb,amsfonts,amsbsy}
\usepackage{algorithmic}
\usepackage{graphicx}
\usepackage{textcomp}
\usepackage{xcolor}

\usepackage[detect-all, binary-units=true]{siunitx}[=v2] % \num{}
\usepackage{multirow} % multirow tables (use like multicolumn)
\usepackage{colortbl} % colored tables (using \cellcolor, \columncolor, \rowcolor)
\usepackage{booktabs} % for \toprule, \middlerule and \bottomrule in Tables
\usepackage{arydshln} % for dashed horizontal lines in tables (\hdashline)
\usepackage{diagbox} % for top left corner of a table with a diagonal line
\usepackage{tikz}
\usepackage{standalone}
\usepackage{pgfplots}
\pgfplotsset{compat=newest} %\pgfplotsset{compat=1.17}
\usepgfplotslibrary{groupplots, polar}
\usepackage[outline]{contour}
\contourlength{1.2pt}

\usepackage[section]{placeins} % for \FloatBarrier
\usepackage[nodisplayskipstretch]{setspace}
\usepackage[caption=false]{subfig} % for subfigures using \subfloat and \(q)quad
\usepackage{import}
\usepackage{dsfont} % for $\mathds{R}^{3}$ or $\mathds{C}^{3}$

\usepackage[hyphens]{url}
\usepackage{acronym}
\newacro{mmWave}[mmWave]{millimeter-wave}
\newacro{THz}{terahertz}
\newacro{RF}{radio frequency}
\newacro{EM}{electromagnetic}
\newacro{RAT}{radio access technology}
\newacro{5GB}{5G and beyond}
\newacro{FR2}{frequency range 2}
\newacro{RAN}{radio access network}
\newacro{CN}{core network}
\newacro{WLAN}{wireless local area network}
\newacro{IEEE}{Institute of Electrical and Electronic Engineers}
\newacro{JCAS}{joint communication and sensing}

\newacro{TX}{transmitter}
\newacro{RX}{receiver}
\newacro{TRX}{transceiver}
\newacro{BS}{base station}
\newacro{UE}{user equipment}

\newacro{SNR}{signal to noise ratio}
\newacro{SINR}{signal to interference plus noise ratio}
\newacro{CSI}{channel state information}
\newacro{RSS}{received signal strength}
\newacro{LOS}{line-of-sight}
\newacro{VLOS}{virtual-line-of-sight}
\newacro{NLOS}{non-line-of-sight}
\newacro{VLOS}{virtual line-of-sight}
\newacro{CIR}{channel impulse response}
\newacro{EIRP}{equivalent isotropically radiated power}
\newacro{UL}{uplink}
\newacro{DL}{downlink}
\newacro{RSRP}{reference signal received power}
\newacro{AOA}{angle of arrival}
\newacro{AOD}{angle of departure}
\newacro{LOB}{line-of-bearing}

\newacro{CDF}{cumulative distribution function}
\newacro{ECDF}{empirical cumulative distribution function}
\newacro{dB}[\si{\decibel}]{decibel}

\newacro{COTS}{commercial-off-the-shelf}
\newacro{RIS}{reconfigurable intelligent surface}
\newacro{XR}{extended reality}

\newacro{UPA}{uniform planar array}
\newacro{HPBW}{half-power beamwidth}
\newacro{SLL}{sidelobe level}
\newacro{FBR}{front-to-back ratio}

\newacro{NUDFT}{non-uniform discrete Fourier transform}

\newacro{B-K}{Beckmann-Kirchhoff}
\newacro{VNA}{vector network analyzer}

\definecolor{tikzsilver}{rgb}{0.7529, 0.7529, 0.7529}
\definecolor{tikzbrown}{rgb}{0.71, 0.4, 0.1}
\definecolor{tikzindigo}{rgb}{0.344117,0,0.69803}
\definecolor{tikzdarkcyan}{rgb}{0.0, 0.75, 0.75}
\newlength\fwidth %for tikz later..
\newlength\fheight %for tikz later..
\newlength\barwidth %for tikz later.
\newcommand\tstrut{\rule{0pt}{2.4ex}} % for use in tables right after \hline for more spacing
\usepackage{hyperref}
\hypersetup{ %hide rectangular boxes around orcid, urls, and FIG/TAB/SEC references
	colorlinks,
	linkcolor={black},
	citecolor={black},
	urlcolor={black}
}
\newcommand\doiurl[1]{%
	\href{https://doi.org/#1}{\nolinkurl{#1}}%
}
\newcommand\mailtocni[1]{%
	\href{mailto:#1@tu-dortmund.de}{\nolinkurl{#1}}%
}
\newcommand\mailtodsv[1]{%
	\href{mailto:#1@uni-due.de}{\nolinkurl{#1}}%
}

\usepackage[
capitalize                  % Always upper case: Section X.Y, Equation (X.Y), ... (use \Cref{lablename})
]{cleveref}                 % grouping and sorting of references
\crefname{table}{Tab.}{Tabs.}
\crefname{section}{Sec.}{Secs.}
\Crefname{section}{Section}{Sections}
\DeclareSIUnit{\dBm}{\deci\bel\of{m}}
\DeclareSIUnit{\dBi}{\deci\bel\of{i}}

\newlength{\origiwspc}
\usepackage[outline]{contour}
\contourlength{1.0pt}

\usepackage{watermark}
\watermark{
	\begin{minipage}{17.9cm}
		\centering
		\footnotesize
		Accepted for presentation in: 2023 6th IEEE International Workshop on Mobile Terahertz Systems (IWMTS), Bonn, Germany, July 2023.
	\end{minipage}
}
\usepackage{eso-pic}% 
\AddToShipoutPictureBG{% Add picture to background of every page
	\AtPageLowerLeft{%
		\raisebox{3\baselineskip}{\makebox[\paperwidth]{\begin{minipage}{21cm}
					\centering
					\footnotesize
					2023 IEEE. Personal use of this material is permitted. Permission from IEEE must be obtained for all other uses,\\
					including reprinting/republishing this material for advertising or promotional purposes, collecting new collected works\\
					for resale or redistribution to servers or lists, or reuse of any copyrighted component of this work in other works.
		\end{minipage}}}%
	}
}

\makeatletter
\newcommand\notsotiny{\@setfontsize\notsotiny\@vipt\@viipt} % between \tiny and \scriptsize
\makeatother
\makeatletter
\newcommand*\sizebetweenfootnotesandscripts{%
	\@setfontsize\sizebetweenfootnotesandscripts{7.5}{9.0}%
}
\makeatother



\begin{document}
\bstctlcite{IEEEexample:BSTcontrol}

\title{
	Phase-based Breathing Rate Monitoring \\
	in Patient Rooms using 6G Terahertz Technology
}
\author{
	\IEEEauthorblockN{
		\textbf{Simon~H{\"a}ger}\textsuperscript{1}\textbf{,}
		\textbf{Akram~Najjar}\textsuperscript{2}\textbf{,}
		\textbf{Caner~Bektas}\textsuperscript{1}\textbf{,}
		\textbf{Dien~Lessy}\textsuperscript{2}\textbf{,}
		\textbf{Mohammed~El-Absi}\textsuperscript{2}\textbf{,}
		\\
		\textbf{Fawad~Sheikh}\textsuperscript{2}\textbf{,}
		\textbf{Stefan~B{\"o}cker}\textsuperscript{1}\textbf{,}
		\textbf{Thomas~Kaiser}\textsuperscript{2}\textbf{,}
		\textbf{and Christian~Wietfeld}\textsuperscript{1}
	}
	% CNI
	\IEEEauthorblockA{
		\textsuperscript{1}Communication Networks Institute (CNI), TU~Dortmund~University, 44227~Dortmund, NRW, Germany
		\\
		E-mail: \{\mailtocni{simon.haeger}, \mailtocni{caner.bektas}, \mailtocni{stefan.boecker}, \mailtocni{christian.wietfeld}\}@tu-dortmund.de
	}
	% DSV
	\IEEEauthorblockA{
		\textsuperscript{2}Institute of Digital Signal Processing, University of Duisburg-Essen, Campus Duisburg, 47057~Duisburg, NRW, Germany
		\\
		E-mail: \{\mailtodsv{akram.najjar}, \mailtodsv{dien.lessy}, \mailtodsv{mohammed.el-absi}, \mailtodsv{fawad.sheikh}, \mailtodsv{thomas.kaiser}\}@uni-due.de
	}
}
\maketitle

%%%%%%%%%%%%%%%%%%%%%%%%%%%%%%%%%%%%%%%%%%%% START OF CONTENT

\noindent
\begin{abstract}
	%
	% Introduction and Problem statement
	%
	The 6G standard aims to be an integral part of the future economy by providing high-performance communication and sensing services.
	At \ac{THz} frequencies, indoor campus networks can offer the highest sensing quality.
	Health monitoring in hospitals is expected to be an application site for these.
	%
	% Results
	%
	This work outlines a monostatic phase-based system for breathing rate monitoring.
	Our feasibility study observes motion measurement accuracy down to the micrometer level.
	However, we also find that the patient's pose needs to be considered for generalized applicability.
	Thus, a solution that leverages multiple propagation paths and beam orientations is proposed.
\end{abstract}


%% Main content
%\acresetall
\section{Introduction}
\label{ch:intro}

\begin{figure}[t!]
	\centering
%	\fbox{
		% trim = left bottom right top
		\includegraphics[clip, trim=4.0cm 4.4cm 4.95cm 4.4cm, width=1.0\columnwidth]{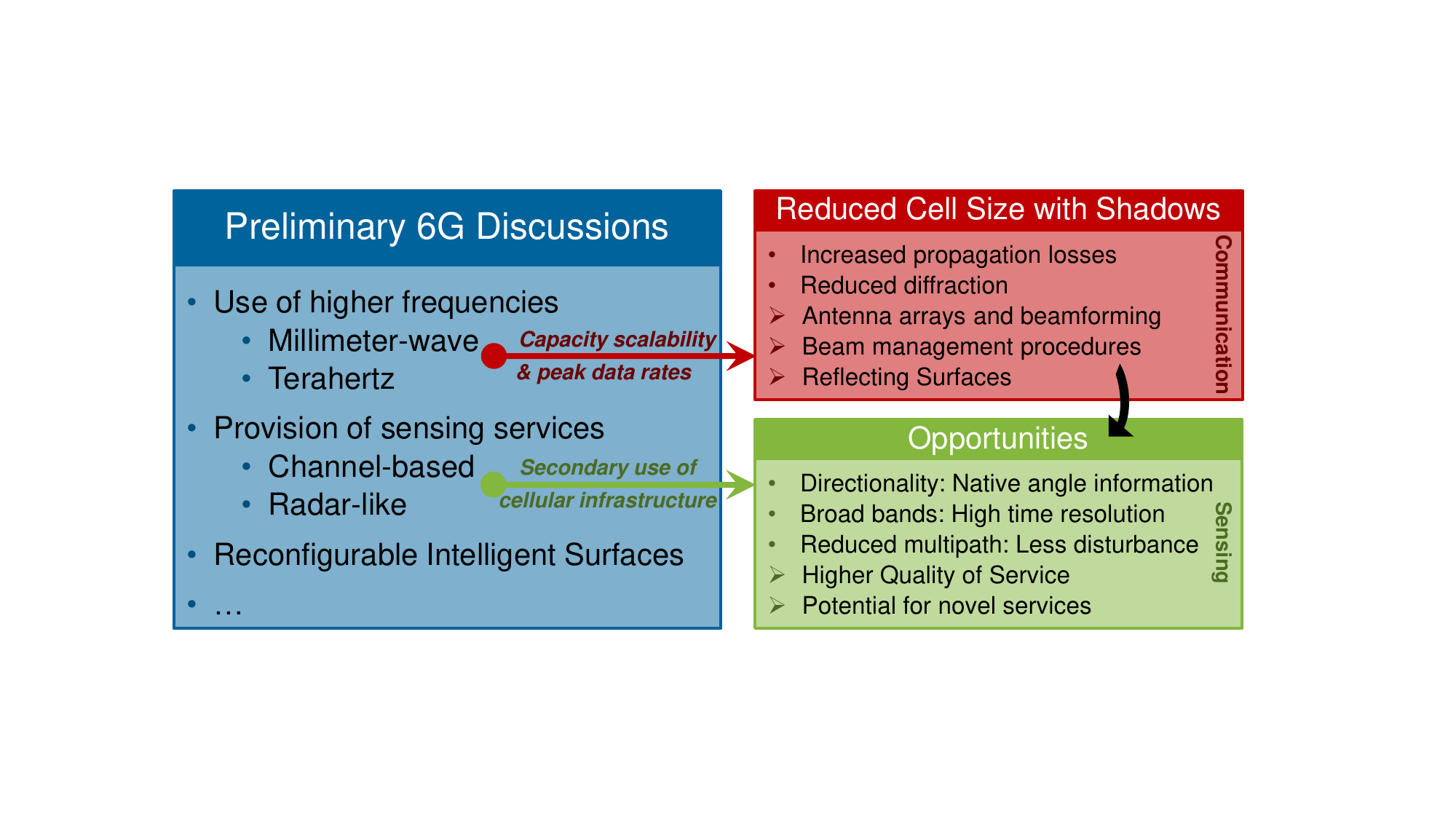}
%	}
	\vspace{-5mm}
	\caption{
		High-performance 6G \ac{THz} campus networks with short-range sensing capabilities: A perfect match for indoor use cases in industry and hospitals. 
	}
	\vspace{-1mm}
	\label{fig:one}
\end{figure}

Mobile radio networks have recently expanded to the \ac{mmWave} domain with the advent of the 5G standard.
The first networks launched operations in urban centers in the US~\cite{AuroraInsights} and test deployments of private campus networks, e.g., connecting manufacturing floors, are up and running~\cite{FranhoferIPT}.
Moreover, the first discussions about what 6G will be have already begun, see \cref{fig:one}.
One of the goals of 6G is to ensure the market success of networks targeting such high frequencies by making it profitable to deploy the technology in the economy, such as in factories and hospitals.

This becomes feasible as the frequencies up to \ac{THz} have risen to the spotlight to reliably provide ultra-high data rates and network capacity at low latency for industrial processes.
Moreover, whereas carriers at \acp{mmWave} and beyond suffer from rather hostile propagation characteristics, they become ever more suitable for indoor sensing purposes in the scope of 6G \ac{JCAS}~\cite{Chaccour/etal/2022}.  
% DSV:
\ac{THz} signals offer superior spatial and temporal resolution than lower bands, such as microwaves, owing to their extremely wide bandwidths and short wavelengths, whereas also exhibiting excellent penetration capabilities compared to optical waves~\cite{Sheikh/etal/2023,fw_2}.
Because \ac{THz} radiation is non-ionizing and non-invasive, it is a safe choice for security screening and medical sensing applications~\cite{Mavis_2023Terahertz}.
As such, machines and processes can be stopped, altered, and (re-)started based on highly accurate sensor information conveniently acquired by the cellular network.
This is of high interest to the health sector for monitoring patients and initializing measures autonomously~\cite{Nayak/Patgiri/2022}.

Communication channels are widely in use to provision sensing services, e.g., the positions of connected \acp{UE}, based on power, time, angle, and phase measurements~\cite{Haeger/etal/2023b}.
% DSV:
The healthcare domain has witnessed the use of \ac{THz} technology for various medical applications, including the characterization of tablet coatings, investigation of drugs, dermatology, oral healthcare, oncology, and medical imaging~\cite{Pawar/etal/2013,Shumaila/etal/2022}. 
There is a wide range of applications in the scope of patient health monitoring, such as human presence detection and tracking, gesture detection, fall detection, tremor monitoring, and tracking of respiration rate~\cite{Ma/etal/2019,Haider/etal/2018}.
6G is expected to introduce radar-like sensing using co-deployed transmit and receive antennas on the \ac{BS} side such that the network autonomously becomes perceptive of the activities in the vicinity, i.e., there is no need for the deployment of \acp{UE}~\cite{Zhang/etal/2021}.
As such, a comparison with the often-considered use cases of breathing rate estimation by tracking the thorax motion~\cite{Tewes/etal/2022} is of interest to characterize the suitability of using \ac{THz} channels in a monostatic fashion.

The remainder of this paper is structured as follows:
\cref{ch:conceptualization} conceptualizes a \ac{THz} phase-based patient monitoring system operating in a monostatic fashion.
Afterward, \cref{ch:meth} describes the ray-tracing-based methodology of this work.
We then evaluate the feasibility of the proposed concept in \cref{ch:eval}.
\cref{ch:concl} concludes this work with an outlook on future work. 
\section{Proposed Patient Monitoring System}
\label{ch:conceptualization}

In this section, we introduce a 6G \ac{THz}-based system for the network-autonomous monitoring of hospital patient rooms for the example use case of breathing rate monitoring as follows. 

\subsection{Breathing Motion Data Acquisition}
\label{ch:conceptualization_A}
During the breathing process, the chest of a typical adult expands by up to \SI{3}{\centi\meter} for a deep breath, whereas a patient's shallow breathing may also result in changes as small as \SI{3}{\milli\meter}~\cite{Tewes/etal/2022}.
These fine-grained motions with magnitude $d$ affect the radio environment, particularly the channel phase of the affected propagation paths~\cite{Tewes/etal/2022, Haeger/etal/2021a}.
Using the complex-valued two-way channel estimate $\hat{h}\left(t\right)$ at time instances $t_{\footnotesize\num{1}}$ (initial state, e.g., lung deflated) and $t_{\footnotesize\num{2}}$ (altered state, e.g., partial inhalation), $t_{\footnotesize\num{2}}>t_{\footnotesize\num{1}}$, the motion can be detected based on measuring a two-way phase difference $\delta\neq0$, as follows~\cite{Haeger/etal/2021a}:
\begin{equation}
	\label{eq:1}
	\delta = \measuredangle\widehat{h}\left(t_{\footnotesize\num{2}}\right) - \measuredangle\widehat{h}\left(t_{\footnotesize\num{1}}\right)
\end{equation}
Leveraging information about the used frequency $f$, the motion magnitude can be estimated using
\begin{equation}
	\label{eq:2}
	\widehat{d} = \frac{
						\delta \cdot c_{\footnotesize\num{0}}
						}{
						2 \cdot 2\pi \cdot f
					   },
\end{equation}
where $c_{\footnotesize\num{0}}$ is the speed of light.
The factor $2\pi$ in \cref{eq:2} transfers the measured two-way phase shift $\delta\in\left(-\pi,\pi\right]$ to a distance in dependence of the wavelength $\lambda$.
Due to the 6G \ac{JCAS}-compatible, monostatic approach using co-deployed transmit and receive antennas per \ac{TRX}, the coefficient of \num{0.5} is introduced to derive a single-way phase change from $\delta$. 

In the previously described process, $\widehat{d}>0$ indicates a shortening of the propagation path between BS and patient, i.e., the chest expands.
Using high signaling rates to attain the channel estimates, the whole inhalation process may be tracked step-by-step.
Similarly, negative motion steps ($\widehat{d}<0$) are measured during the deflation process of the lung.
Continuing the process indefinitely, a time series of fine motion steps is captured by the 6G \ac{TRX}, thus allowing for the reconstruction of the overall chest movement as well as the breathing rate.

\begin{figure}[t!]
	\centering
	%	\fbox{
	% trim = left bottom right top
	\includegraphics[clip, trim=2.175cm 3.45cm 2.175cm 3.25cm, width=1.0\columnwidth]{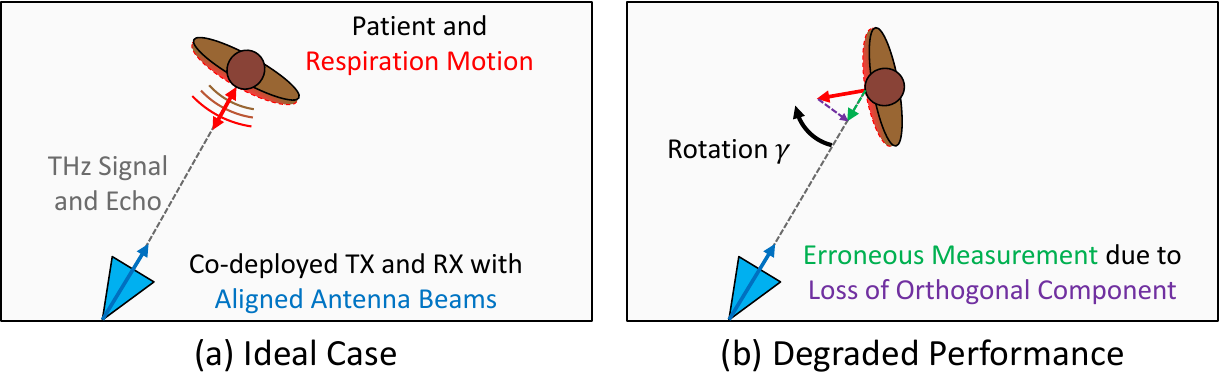}
	%	}
	\vspace{-5mm}
	\caption{
%		Patient room from a bird's eye view.
		Top view of the patient room.
		Phase-based respiration monitoring cannot detect motion components that are orthogonal to the propagation path.
	}
	\label{fig:rotation}
	\vspace{-2mm}
\end{figure}

\subsection{Conquering Limitations by the Patient Pose}
\label{ch:conceptualization_B}
One of the implicit assumptions in \cref{ch:conceptualization_A} is that the thorax motion matches the propagation direction, i.e., the Poynting vector of the \ac{EM} wave in 3D space, as illustrated in {\cref{fig:rotation}~(a)}.
However, if the user is rotated by $\gamma$, the breathing motion of the patient contains an orthogonal component that cannot be measured~\cite{Haeger/etal/2021a}, see {\cref{fig:rotation}~(b)}.
As such, the motion magnitude measurement suffers from a systematic error such that
\begin{equation}
	\label{eq:3}
	\widehat{d} = d\cdot\cos\left(\gamma\right).
\end{equation}
As a consequence, at $\left|\gamma\right|\geq\SI{90}{\degree}$ rotation angles no motion can be observed because the motion is either fully orthogonal to the propagation path or the patient's back is observed instead of the front side.
The beam orientation-aided approach described in \cite{Haeger/etal/2021a} has remedied this problem, however, assuming the receiver is mounted at the chest of the patient.

\begin{figure}[b!]
	\vspace{-2mm}
	\centering
%	\fbox{
		% trim = left bottom right top
		\includegraphics[clip, trim=5.8cm 3.8cm 4.2cm 3.75cm, width=1.0\columnwidth]{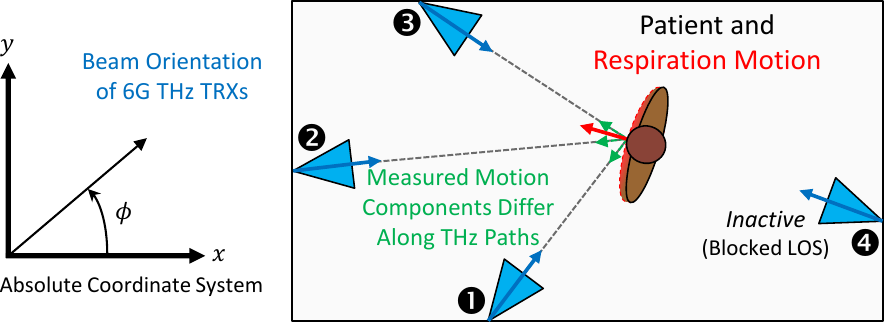}
%		}
	\vspace{-5mm}
	\caption{
		Reconstructing correct breathing motion using four transceivers that monitor the whole patient room with their well-aligned antenna beams.
		The antenna orientations $\left(\hat{\phi}_{\footnotesize{n}}, \hat{\theta}_{\footnotesize{n}}\right)$ (blue arrows) and measured motion components $\hat{d}_{\footnotesize{n}}$ (green arrows) are used to reconstruct the original respiration motion (red arrow).
		A sufficiently larger number of nodes guarantees service availability. % despite potential propagation path blockages.
	}
	\label{fig:respiration_reconstruction}
	\vspace{-1mm}
\end{figure}

In the following section, we evolve the technique from \cite{Haeger/etal/2021a} to the considered monostatic sensing scenario using multiple infrastructure nodes as shown in \cref{fig:respiration_reconstruction}.
Nodes $n$ with $n=1,\dots,N$ have measured the motions $\widehat{d}_{\footnotesize{n}}$ using the antenna beam orientation tuples $\left(\widehat{\phi}_{\footnotesize{n}}, \widehat{\theta}_{\footnotesize{n}}\right)$, which are carefully aligned to the patient's torso.
The definition of azimuth angle $\phi$ is depicted in \cref{fig:respiration_reconstruction}, whereas the elevation angle $\theta$ is defined from the x/y-plane towards the z-axis.
As such, the beam orientation in space is as follows.
\begin{equation}
	\label{eq:4}
	\vec{e}\left(\phi,\theta\right) =
	\left[
		\cos\left(\phi\right)\cos\left(\theta\right), \sin\left(\phi\right)\cos\left(\theta\right),
		\sin\left(\theta\right)
	\right]^{T}
%	\begin{bmatrix}
%		\cos\left(\phi\right)\cos\left(\theta\right) \\
%		\sin\left(\phi\right)\cos\left(\theta\right) \\
%		\sin\left(\theta\right)
%	\end{bmatrix}
\end{equation}
Patient motion $\widehat{d}_{\footnotesize{n}}$ can thus be matched to the direction vector
\begin{equation}
	\label{eq:5}
	\vec{r}_{\footnotesize{n}} = -\vec{e}\left(\widehat{\phi}_{\footnotesize{n}},\widehat{\theta}_{\footnotesize{n}}\right),
\end{equation}
such that motion components along the two mutually orthogonal vectors $\vec{e}_{\footnotesize{A,n}}, \vec{e}_{\footnotesize{B,n}}\hspace{0.5mm}\bot\hspace{0.5mm}\vec{r}_{\footnotesize{n}}$ have been lost.
Therefore, the idea is to recover their coefficients, such that, e.g.,
\begin{equation}
	\label{eq:6}
	\vec{d} = \widehat{d}_{\footnotesize\num{1}} \cdot\vec{r}_{\footnotesize\num{1}} +
	\widehat{c}_{1} \cdot\vec{e}_{\footnotesize{A,1}} + 
	\widehat{c}_{2} \cdot\vec{e}_{\footnotesize{B,1}}.
\end{equation}
They are recovered by solving the linear equation system of $\vec{b} = A \cdot \vec{c}$.
Vector $\vec{c}\in\mathds{R}^{2N\times1}$ shall be determined from matrix $A$ and vector $\vec{b}$ as follows.
\begin{equation}
	\label{eq:7}
	\vec{b} =
	\begin{bmatrix}
		\widehat{d}_{\footnotesize\num{2}} \cdot \vec{r}_{\footnotesize\num{2}} - \widehat{d}_{\footnotesize\num{1}} \cdot \vec{r}_{\footnotesize\num{1}} \\
		... \\
		\widehat{d}_{\footnotesize{n}} \cdot \vec{r}_{\footnotesize{n}} - \widehat{d}_{\footnotesize\num{1}} \cdot \vec{r}_{\footnotesize\num{1}}
	\end{bmatrix}
\end{equation}
\begin{equation}
	\label{eq:8}
	A = 
	\begin{pmatrix} 
		\vec{e}_{\footnotesize{A,1}}, & \vec{e}_{\footnotesize{B,1}}, &
		-\vec{e}_{\footnotesize{A,2}}, & -\vec{e}_{\footnotesize{B,2}}, &
		0_{3x1}, & 0_{3x1} \\
		\vec{e}_{\footnotesize{A,1}}, & \vec{e}_{\footnotesize{B,1}}, &
		0_{3x1}, & 0_{3x1} , &
		-\vec{e}_{\footnotesize{A,3}}, & -\vec{e}_{\footnotesize{B,3}}
	\end{pmatrix}
\end{equation}
Note that $\vec{b} \in \mathds{R}^{3\left(N-1\right)\times1}$ and $A \in \mathds{R}^{3\left(N-1\right)\times2N}$.
For brevity, \cref{eq:8} defines $A$ for the case of $N=3$.
Note that $N$ may not be smaller for a unique solution.
At last, the overall breathing motion is reconstructed using all $2N$ elements contained in $\vec{c}$:
\begin{equation}
	\label{eq:9}
	\vec{d} = \underset{u=1,\dots,N}{\text{mean}} \left\{
	\widehat{d}_{\footnotesize{u}} \cdot\vec{r}_{\footnotesize{u}} +
	\widehat{c}_{\footnotesize{2u-1}} \cdot\vec{e}_{\footnotesize{A,u}} + 
	\widehat{c}_{\footnotesize{2u}} \cdot\vec{e}_{\footnotesize{B,u}} \right\}.
\end{equation}
Thus, the sought-after breathing motion magnitude can be extracted using the Euclidean norm, i.e., by $\widehat{d}=\left\lVert\vec{d}\hspace{0.5mm}\right\rVert_{\footnotesize\num{2}}$.
As such, we have shown with \cref{eq:7,eq:8,eq:9} how angle information from $N$ nodes (\cref{eq:2}) along with the phase changes of the corresponding channels (\cref{eq:5}) can be used to measure the fine motion irrespective of the patient's pose, i.e., position and orientation in the patient room.

\subsection{Network Planning for Optimized THz Indoor Deployments}
\label{ch:conceptualization_C}
The above-proposed patient monitoring system in \cref{ch:conceptualization_A,ch:conceptualization_B} relies on $N$ co-deployed \acp{TRX}, as depicted in \cref{fig:respiration_reconstruction}.
However, depending on the room dimension, furnishing, and the position and rotation angle of the patient, they have to be mounted intelligently at suitable positions such that at least three paths exhibit \ac{LOS} modality.
In our earlier work~\cite{Bektas/etal/2021a}, automatic network planning based on machine learning was developed and evaluated for sub-\SI{6}{\giga\hertz} communication networks.
There, all possible antenna positions were simulated beforehand via ray-tracing for subsequent use of clustering methods to identify the best local results.
Afterward, they are combined into a network plan which suits the application requirements, such as the need for $\geq\hspace{-0.5mm}\num{3}$ distinct propagation paths.
This needs to be explored in future work.

\begin{figure}[t!]
	\centering
%	\fbox{
	% trim = left bottom right top
	\includegraphics[clip, trim=1.675cm 2.275cm 1.225cm 1.05cm, width=1.0\columnwidth]{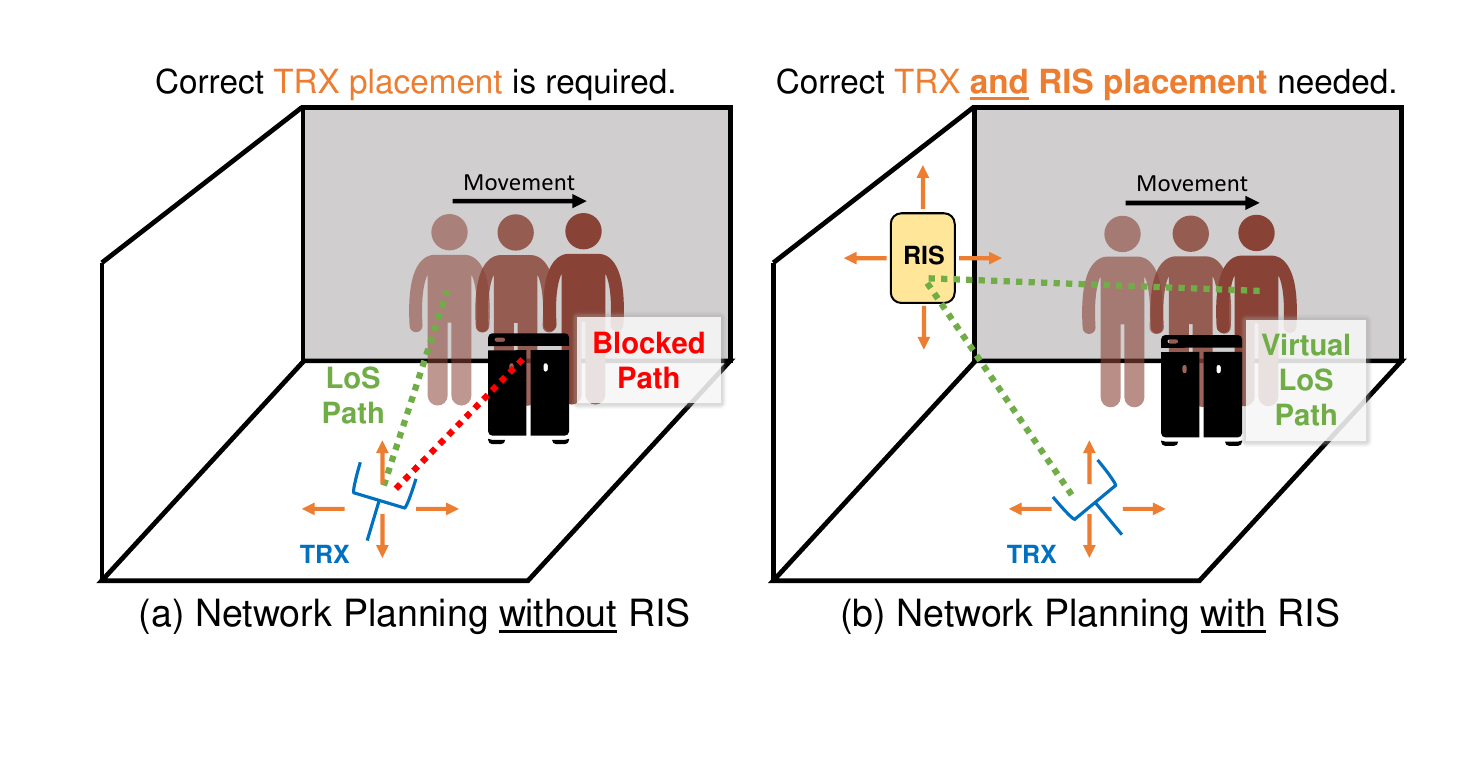}
%		}
	\vspace{-5mm}
	\caption{
		Introducing the idea of using \acp{RIS} instead of multiple transceivers.
		Schematic insight into the ensuing \ac{RIS}-dependent network planning overhead to provide sufficient \ac{THz} link availability at all possible patient positions.
	}
	\label{fig:planning}
	\vspace{-2mm}
\end{figure}

Moreover, the requirement for $N$ \acp{TRX} is not optimal due to costs. 
Considering the emerging topic of \acp{RIS}, it could be reduced to just one if $N-1$ \acp{RIS} are employed to provide \ac{VLOS} paths along well-selected reflection beams.
Therefore, multipath propagation could be exploited as proposed in \cite{Haeger/etal/2022a}.
However, this adds a new dimension to the pre-deployment network planning, as illustrated in \cref{fig:planning}.
Whereas algorithms for the placement of \acp{RIS} have already been proposed in the literature, e.g., in \cite{ElAbsi/etal/2022}, they have to be altered for the previously described requirements.
Particularly the planning overhead needs to be considered due to the dependency of the \ac{RIS} placement on the \ac{TRX} placement.
Whereas the overhead for traditional network planning is in $\mathcal{O}\hspace{-0.5mm}\left(M\right)$~\cite{Bektas/etal/2021a}, with $M$ being the number of considered mounting positions, the planning of \ac{RIS}-assisted deployments is in $\mathcal{O}\hspace{-0.5mm}\left(M^2\right)$.
In conclusion, previously possible automated network planning methods have to be extended by a pre-processing of possible position combinations for antenna and \acp{RIS} to limit the number of required simulations.
\section{Methodology}
\label{ch:meth}

This section details the methodology of this work.
\cref{ch:meth_A} describes the considered indoor scenario and gives details on the ray-tracing simulations.
Afterward, \cref{ch:meth_C} gives additional details on the signal processing exceeding the scope of \cref{ch:conceptualization}.
\cref{ch:meth_D} concludes this section with details on the evaluation in \cref{ch:eval}.

\subsection{Patient Room Scenario and Terahertz Ray-tracing}
\label{ch:meth_A}

\begin{figure}[b!]
	\vspace{-2mm}
	\centering
%		\fbox{
			% trim = left bottom right top
			\includegraphics[clip, trim=0cm 0cm 7.1cm 1.15cm, width=0.768175\columnwidth]{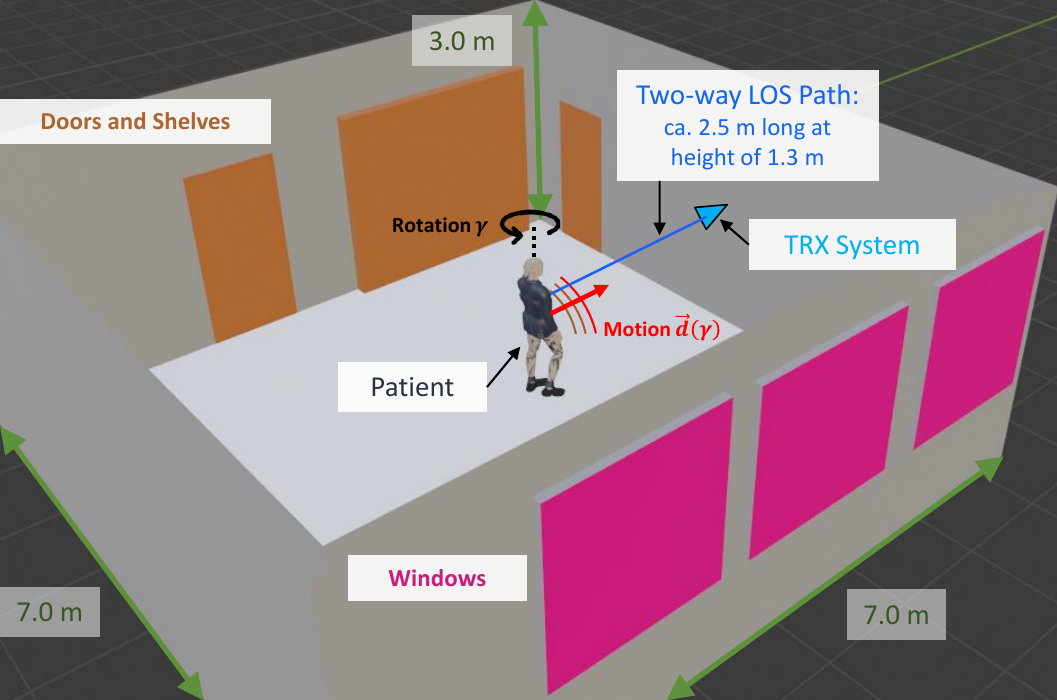}
%		}
	\vspace{-1mm}
	\caption{
		Patient room scenario under consideration in this work.
	}
	\label{fig:scenario}
\end{figure}
\begin{figure}[b!]
	\vspace{-2mm}
	\centering
	%	\fbox{
	% trim = left bottom right top
	\includegraphics[clip, trim=0.275cm 0.290cm 14.675cm 10.205cm, width=0.768175\columnwidth]{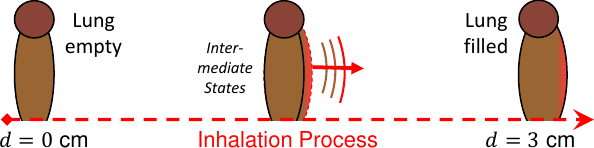}
	%	}
	\vspace{-1mm}
	\caption{
		Sketch of considered breathing motion with uniform motion steps between simulation-based channel estimates from first to last index.
	}
	\label{fig:breathinganimation}
	\vspace{-1mm}
\end{figure}

This work considers the $\num{7}\times\num{7}\times\SI{3}{\meter}$ indoor scenario depicted in \cref{fig:scenario}.
For brevity, detailed descriptions of objects such as doors, cupboard, and glass windows are omitted as they can be found in~\cite{Sheikh/etal/2021}.
The \ac{JCAS} \ac{TRX} is mounted at a height of \SI{1.3}{\meter}.
The patient, and thus, the breathing motion with magnitude $d$, is initially perfectly aligned to the \ac{LOS} propagation path (using $\gamma=\SI{0}{\degree}$) with a length of \SI{2.55}{\meter} to the patient body's center point.
In adapted scenarios the user is rotated such that the breathing motion is partially orthogonal instead of fully aligned to the \ac{LOS} path.
The angles $\gamma \in \left\{\SI{22.5}{\degree}, \SI{45.0}{\degree}, \SI{60.0}{\degree}, \SI{90.0}{\degree}\right\}$ are considered.
To achieve perfect scattering in this work, the human body of the patient is attributed with perfect electric conductor-like properties.
Otherwise, the performance of the system could be impaired.

The patient reflection-based channels between co-deployed \ac{TX} and \ac{RX} are simulated for the above described scenario as follows.
Horn antennas with approx. \SI{25}{\dBi} gain and \SI{10}{\degree} E-plane \ac{HPBW} are employed based on \ac{EM} simulations with {CST}~\cite{CST}.
They are vertically polarized to optimize scattering effects and reduce scattering losses.
Using a transmit power of \SI{-10}{\dBm}, the patient reflection-based multipath channel is captured along the breathing inhalation motion of up to \SI{3}{\centi\meter} shown in \cref{fig:breathinganimation,fig:scenario}.
At \SI{100}{\giga\hertz} the patient is moved in steps of \SI{0.5}{\milli\meter} using the commercial ray-tracer {Wireless Insite}~\cite{Remcom}.
Moreover, we also leverage channel data at \SI{300}{\giga\hertz} in steps of about \SI{166.7}{\micro\meter} using a \ac{THz}-specific commercial ray-tracer~\cite{TMTC}.
In both cases, the requirement for motion steps smaller than $\num{0.25}\lambda$ between channel samples, i.e., a high sampling rate, is enforced to avoid \ac{EM} wave-based phase ambiguities.

\begin{table}[t!]%
	\vspace{-1mm}
	\caption{
		Summary of simulated wireless channels~\cite{Remcom,TMTC}.
	}
	\vspace{-1mm}
	\label{tab:scenario}
	%	\resizebox{\columnwidth}{!}{
	\sizebetweenfootnotesandscripts
	\centering
	\begin{tabular}{l@{\hspace{1.5mm}}l@{\hspace{1.5mm}}l}
		\hline
		\rowcolor{lightgray}
		\tstrut \textbf{Carrier} & \tstrut \textbf{Parameter} & \tstrut \textbf{Value/Range} \\
		
		%%%
		\hline 
		\tstrut \SI{100}{\giga\hertz}
		& Motion Trajectory
		& $d = \left[\SI{0}{\centi\meter}:\SI{500}{\micro\meter}:\SI{3}{\centi\meter}\right]$ \\
		& Rotation Angles
		&  $\gamma \in \left\{\SI{0.0}{\degree}, \SI{22.5}{\degree}, \SI{45.0}{\degree}, \SI{60.0}{\degree}, \SI{90.0}{\degree}\right\}$\\
		& Number of Simulations 
		& $\num{61}\cdot\num{5}=\num{305}$ \\
		
		%%%
		\hline 
		\tstrut \SI{300}{\giga\hertz}
		& Motion Trajectory 
		&$d = \left[\SI{0}{\centi\meter}:\SI{167}{\milli\meter}:\SI{3}{\centi\meter}\right]$ \\
		& Rotation Angles 
		& $\gamma \in \left\{\SI{0.0}{\degree}, \SI{22.5}{\degree}, \SI{45.0}{\degree}, \SI{60.0}{\degree}, \SI{90.0}{\degree}\right\}$\\
		& Number of Simulations 
		& $\num{181}\cdot\num{5}=\num{905}$ \\
		
		\hline
	\end{tabular}
	%	}
	\vspace{-1mm}
\end{table} 
\begin{table}[t!]%
	\vspace{-1mm}
	\caption{
		Material parameters used in \ac{THz} ray-tracing~\cite{fsheikh2}.
	}
	\vspace{-1mm}
	\label{tab:simdetails}
	%	\resizebox{\columnwidth}{!}{
	\sizebetweenfootnotesandscripts
	\centering
	\begin{tabular}{l@{\hspace{3mm}}l@{\hspace{3mm}}c@{\hspace{3mm}}c@{\hspace{3mm}}c@{\hspace{3mm}}c}
		\hline
		\rowcolor{lightgray}
		\tstrut \textbf{Material} &
		\tstrut \textbf{Type} &
		\tstrut $\boldsymbol {\tilde{\epsilon}_{r}}$ &
		\tstrut $\boldsymbol {\tilde{\tilde{\epsilon}}_{r}}$ &
		\tstrut $\boldsymbol {\ell_{cr}}$ &
		\tstrut $\boldsymbol {\sigma_h}$ \\ 

		\hline\tstrut
		Plaster 	& Rough		& 3.691 & 0.217	& \SI{1.50}{\milli\meter} & \SI{0.15}{\milli\meter} \\
		\hline\tstrut
		PVC 		& Smooth 	& 2.788	& 0.069 & - & \SI{0.00}{\milli\meter} \\
		\hline\tstrut
		Wood 		& Smooth 	& 1.734 & 0.073 & -  & \SI{0.00}{\milli\meter}  \\
		\hline\tstrut
		Glass 		& Smooth 	& 6.656 & 0.539 & - & \SI{0.00}{\milli\meter} \\
		
		\hline
	\end{tabular}
	%	}
	\vspace{-2mm}
\end{table}

We now give some further simulation details, in particular, for the \SI{300}{\giga\hertz} simulations.
\cref{tab:simdetails} enlists the empirical parameters of the object materials employed in our modeling approach.
As only the walls and ceiling are made of rough plaster, the statistical roughness parameters, such as the standard deviation of height ($\sigma_h$) and surface correlation length ($\ell_{cr}$), are also provided. The \ac{B-K} approach is utilized to approximate the attenuation caused by surface roughness in a specular direction of reflection, as well as diffuse scattering in non-specular directions.
It is implemented using a tile size of \num{10}$\ell_{cr}$ and $\num{20}\times\num{20}$ tiles, as in~\cite{fsheikh2}.
The simulations further take into account state-of-the-art \SI{300}{GHz} \ac{VNA}~\cite{fw_2} based channel measurements for calibration of the ray-tracer~\cite{TMTC}.
Abiding the {WR-3.4} extenders' \SI{105}{\dB}-restricted peak dynamic range, multipath components incurring higher losses are ignored.

\subsection{Additional Processing Methodology}
\label{ch:meth_C}

The result of the ray-tracing simulations are complex-valued \acp{CIR} for each state along the patient's motion.
To extract and compare the channel phases over time, the \ac{CIR} vector $\vec{h}\left(\tau\right)|_{\footnotesize{t}}$ acquired at each time instance $t$ is first transformed to the frequency-domain using the \ac{NUDFT}, thus allowing for a similar processing as in \cite{Haeger/etal/2021a}.
Accordingly, a phase unwrapping operation is conducted for each frequency-based time series.
Last, the mean channel phase over all frequencies is estimated per time instance resulting in one channel phase estimate per \ac{CIR}.
Note that the process could be refined if the \ac{TRX} leverages information about the baseline scenario, i.e., the patient room without patient inside.

\subsection{Evaluation Methodology}
\label{ch:meth_D}

This work will first investigate the feasibility of the monostatic breathing rate monitoring in \cref{ch:eval_A} for the ideal case as sketched in {\cref{fig:rotation}~(a)}.
We will look into the measured motion steps $\hat{d}$ between consecutive time instances along an inhalation process, thus representing any respiration pattern, and compare to the expected value. 
This applies to the data provided by both ray-tracers, i.e., at \SI{100}{\giga\hertz} with a step size of \SI{0.5}{\milli\meter} and at \SI{300}{\giga\hertz} in steps of about \SI{166.7}{\micro\meter} ($\lambda/\num{6}$).

Afterward, in \cref{ch:eval_B} we consider $\gamma$-rotated patients which are expected to impact the practical performance, thus motivating our proposed reconstruction technique.
Hence, we investigate the reconstructed overall chest motion which arises from continuous integration of the detected motion steps $\hat{d}$.
The ensuing trajectory of the patient's chest during the inhalation process is evaluated for the considered four rotation angles of up to \SI{90.0}{\degree}, cf.~\cref{fig:scenario} for the rotation definition, and compared to the performance for the non-rotated patient.
\section{Evaluation}
\label{ch:eval}

This section first studies the feasibility of channel phase-based patient breathing rate monitoring using a 6G \ac{THz} \ac{TRX} in a monostatic fashion as in the scope of \ac{JCAS}.
Afterward, a sensitivity analysis of the impact of the user rotation is conducted as described in \cref{ch:meth_D}.

\subsection{Feasibility Study of THz Phase-based Patient Monitoring}
\label{ch:eval_A}

\begin{figure}[t!]
	\setlength\fwidth{0.915\linewidth}
	\setlength\fheight{0.275\fwidth}
	\centering

% Thesis colors:
\definecolor{mycolor1}{rgb}{0.894118,0.101961,0.109804}% red
\definecolor{mycolor2}{rgb}{0.344117,0,0.69803}% indigo
\definecolor{mycolor3}{rgb}{0.71, 0.4, 0.1}% brown
\definecolor{mycolor4}{rgb}{0.00000,0.4400,0.85100}% blue
\definecolor{mycolor5}{rgb}{0.301961,0.686275,0.290196}% green
\definecolor{mycolor6}{rgb}{0.25, 0.25, 0.28}% gray/black

\begin{tikzpicture}

\begin{axis}[%
	width=0.951\fwidth,
	height=\fheight,
	at={(0\fwidth,0\fheight)},
	scale only axis,
	xmin=-5,
	xmax=185,
	xlabel style={font={\footnotesize \color{white!15!black}}, yshift=0.75mm},
	xlabel={Channel Estimate Index},
	ymin=165.416667,
	ymax=167.916667,
	ylabel style={font={\footnotesize \color{white!15!black}}},
	ylabel={Motion Steps $\left[\si{\micro\meter}\right]$},
	axis background/.style={fill=white},
	xmajorgrids,
	ymajorgrids,
	legend style={at={(0.5,1.05)}, anchor=south, legend cell align=left, align=left, draw=white!15!black,font=\footnotesize, row sep=-0.02cm, inner xsep=2pt, inner ysep=0pt},
	legend columns=2,
	minor x tick num=3,
	minor y tick num=4,
	ticklabel style={font=\scriptsize},
	]
	%% Annotation
	\node (source) at (axis cs:95,166.55){};
	\node (destination) at (axis cs:95,167.775){};
	\draw[->|, line width=1pt, color=mycolor6] (source)--(destination);	
	\node[anchor=north east, xshift=0.5mm, yshift=-0.5mm] at (destination) {
		\scriptsize
		\contour{white}{
			\textcolor{mycolor6}{
				Sub-\SI{1}{\micro\meter} Accuracy
			}
		}
	};
	
	%% LINES
	\addplot [color=black, line width=1.0pt]
	table[row sep=crcr]{%
		1	166.666666666657\\
		180	166.666666666657\\
	};
	\addlegendentry{Ground Truth}
	
	\addplot [color=mycolor1, line width=0.75pt]
	table[row sep=crcr]{%
		1	166.189122976789\\
		2	166.695997645832\\
		7	166.648097431533\\
		8	166.269338625824\\
		9	166.677254588791\\
		10	166.293219805169\\
		11	166.699882389651\\
		15	166.739143138756\\
		16	166.261976600936\\
		17	166.670874803174\\
		18	166.780935917245\\
		19	166.190790076371\\
		20	167.20758242252\\
		21	166.717218258214\\
		22	166.727485744161\\
		23	166.138172196952\\
		24	166.757606362467\\
		25	166.259066779304\\
		26	166.675320453805\\
		27	166.686140603844\\
		28	166.301543230389\\
		30	167.111104333031\\
		31	166.243100236045\\
		32	167.239771968866\\
		33	166.252957793404\\
		34	166.170945070235\\
		35	166.778995731562\\
		37	166.702787872538\\
		38	166.713608715952\\
		39	166.228245745486\\
		40	166.730680087142\\
		41	166.648521911071\\
		42	166.271147873402\\
		43	167.161369113188\\
		44	166.286833898539\\
		45	166.694876248755\\
		46	166.707224037197\\
		47	166.221308635158\\
		48	167.219367217115\\
		49	166.15289088125\\
		50	166.745265865842\\
		51	166.66587709679\\
		52	166.278506091004\\
		53	167.185651007486\\
		54	166.202295061112\\
		55	166.703890333423\\
		56	166.730263316481\\
		57	166.236857999536\\
		58	166.739434956185\\
		59	166.657413011592\\
		60	166.272398148323\\
		61	166.680583967491\\
		62	166.792028819174\\
		64	166.714964057178\\
		65	166.224656070312\\
		66	166.640195074109\\
		67	166.758879891034\\
		68	166.257397981724\\
		69	167.172226852898\\
		70	166.684401222739\\
		71	166.693390954322\\
		72	166.211595433992\\
		73	166.712774861907\\
		74	166.739157032652\\
		75	166.245195450045\\
		76	166.748459777722\\
		77	166.666151576351\\
		78	167.183143671167\\
		79	166.284877365968\\
		80	166.701393440121\\
		81	166.713470741797\\
		82	166.226177489534\\
		83	166.231596086618\\
		84	167.145822148171\\
		85	166.267950567715\\
		86	167.169967104599\\
		87	166.277101640901\\
		88	166.196464318546\\
		89	167.203966484834\\
		90	166.219652268978\\
		91	166.721672552001\\
		93	166.751923806885\\
		94	166.756082435313\\
		95	166.177728574894\\
		96	166.687095860198\\
		98	166.709870657968\\
		99	166.323890626671\\
		100	166.633400883484\\
		101	166.238947979821\\
		102	167.654903369112\\
		103	165.77745058774\\
		104	167.172455045289\\
		105	166.691542729609\\
		106	165.805913750319\\
		107	167.213271777371\\
		108	166.227431979877\\
		109	167.130552599332\\
		110	166.25726259764\\
		111	166.761628921648\\
		112	166.166485616429\\
		113	166.784948261636\\
		115	166.70834751659\\
		116	166.719174687103\\
		117	166.231314347414\\
		118	167.142207601074\\
		119	165.748443410365\\
		120	166.766068341957\\
		121	167.18342248555\\
		123	166.200918216333\\
		124	166.21244303667\\
		125	167.223546000855\\
		127	166.240743120767\\
		128	167.165237583362\\
		129	166.270996025048\\
		130	166.674474170577\\
		131	166.193423716209\\
		132	166.705710684681\\
		135	166.739701602864\\
		136	166.251981458657\\
		137	167.155524057932\\
		138	166.275580459313\\
		139	166.185930334706\\
		140	166.697521746215\\
		143	166.733459057752\\
		144	166.2439300041\\
		145	166.649206853663\\
		146	166.774542200187\\
		147	166.171491654239\\
		148	167.1914863702\\
		149	166.301686381757\\
		150	166.715007293277\\
		151	166.624782107732\\
		153	166.748719672938\\
		154	166.259759095112\\
		155	167.165942957212\\
		156	165.782460077395\\
		157	166.695719394255\\
		158	167.106957277326\\
		160	166.229082030441\\
		161	166.742336081541\\
		162	166.24532214837\\
		163	167.165527039629\\
		164	166.283086644256\\
		165	166.681005456184\\
		166	166.700160889008\\
		167	166.210348428649\\
		168	167.225728576256\\
		169	166.234444768065\\
		170	166.746819076991\\
		171	166.157624850769\\
		172	166.761723974673\\
		173	166.681137146696\\
		174	166.288304077284\\
		175	167.105563762846\\
		176	166.314440002111\\
		177	167.128456025097\\
		178	166.238383332342\\
		179	166.750397143261\\
		180	166.662683255737\\
	};
	\addlegendentry{Measurement}
		
\coordinate (annotate_empty) at (0,166.666667);
\coordinate (annotate_full) at (180,166.666667);		
				
\end{axis}

\node[align=left, font=\notsotiny, yshift=-16mm, xshift=3.9mm] at (annotate_empty) {(Lung Empty)};
\node[align=right, font=\notsotiny, yshift=-16mm, xshift=-2.7mm] at (annotate_full) {(Lung Full)};

\end{tikzpicture}%
	\vspace{-7mm}
	\caption{
		Individually measured motion steps between consecutive channel estimates at \SI{300}{\giga\hertz} matches the underlying patient chest motion as desired. 
	}
	\label{fig:steps_ideal_300GHz}
\end{figure}
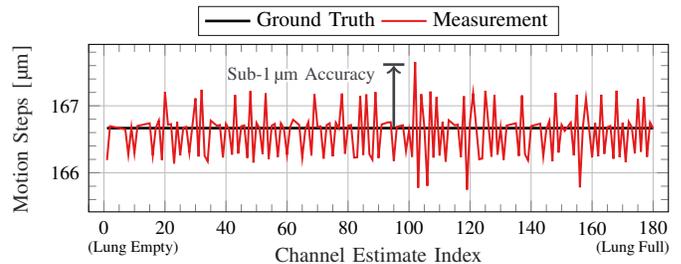
\begin{figure}[t!]
	\vspace{-3mm}
	\setlength\fwidth{0.905\linewidth}
	\setlength\fheight{0.275\fwidth}
	\centering

% Thesis colors:
\definecolor{mycolor1}{rgb}{0.894118,0.101961,0.109804}% red
\definecolor{mycolor2}{rgb}{0.344117,0,0.69803}% indigo
\definecolor{mycolor3}{rgb}{0.71, 0.4, 0.1}% brown
\definecolor{mycolor4}{rgb}{0.00000,0.4400,0.85100}% blue
\definecolor{mycolor5}{rgb}{0.301961,0.686275,0.290196}% green
\definecolor{mycolor6}{rgb}{0.25, 0.25, 0.28}% gray/black

\begin{tikzpicture}

\begin{axis}[%
	width=0.951\fwidth,
	height=\fheight,
	at={(0\fwidth,0\fheight)},
	scale only axis,
	xmin=-1,
	xmax=61.6,
	xlabel style={font={\footnotesize \color{white!15!black}}, yshift=0.75mm},
	xlabel={Channel Estimate Index},
	ymin=0.4925,
	ymax=0.5175,
	ylabel style={font={\footnotesize \color{white!15!black}}},
	ylabel={Motion Steps $\left[\si{\milli\meter}\right]$},
	axis background/.style={fill=white},
	xmajorgrids,
	ymajorgrids,
	legend style={at={(0.974,0.8975)}, anchor=north east, legend cell align=left, align=left, draw=white!15!black,font=\footnotesize, row sep=-0.02cm, inner xsep=2pt, inner ysep=0pt},
	legend columns=2,
	minor tick num=4,
	ticklabel style={font=\scriptsize},
	]
	
	%% Annotation
	\node (source) at (axis cs:6,0.49875){};
	\node (destination) at (axis cs:6,0.515){};
	\draw[->|, line width=1pt, color=mycolor6] (source)--(destination);	
	\node[anchor=north west, xshift=-0.75mm, yshift=-8.5mm] at (destination) {
		\scriptsize
		\contour{white}{
			\textcolor{mycolor6}{
				Incurred Error $\ll$ Motion Step Size
			}
		}
	};
	
	\addplot [color=black, line width=1.0pt]
	table[row sep=crcr]{%
		1	0.5\\
		60	0.5\\
	};
	\addlegendentry{Ground Truth}
	
	\addplot [color=mycolor1, line width=1.0pt]
	table[row sep=crcr]{%
		1	0.496425303714958\\
		2	0.496420522138955\\
		3	0.5139547558502\\
		4	0.496455451427607\\
		59	0.496283328319706\\
		60	0.495027103083551\\
	};
	\addlegendentry{Measurement}

\coordinate (annotate_empty) at (0,0.5);
\coordinate (annotate_full) at (60,0.5);	

\end{axis}

\node[align=left, font=\notsotiny, yshift=-11.5mm, xshift=4.65mm] at (annotate_empty) {(Lung Empty)};
\node[align=right, font=\notsotiny, yshift=-11.5mm, xshift=-2.7mm] at (annotate_full) {(Lung Full)};

\end{tikzpicture}%
	\vspace{-7mm}
	\caption{
		Measured motion steps at \SI{100}{\giga\hertz} approaches ground truth but incurred maximum absolute error is higher than at \SI{300}{\giga\hertz}.	
	}
	\label{fig:steps_ideal_100GHz}
	\vspace{-3mm}	
\end{figure}
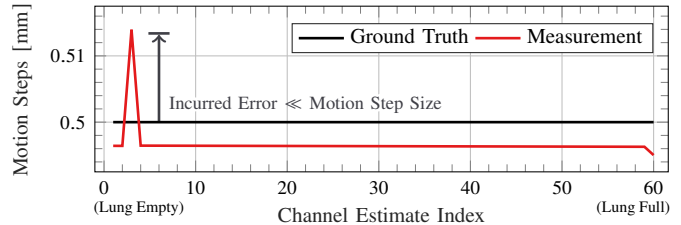 

We first consider the measured motion steps between consecutive channel estimates for the \SI{300}{\giga\hertz} simulation data, as depicted in \cref{fig:steps_ideal_300GHz}.
The motion steps observed oscillate around the underlying actual patient motion sampled about every \SI{166.7}{\micro\meter}.
The absolute error does not exceed \SI{1}{\micro\meter}.
Therefore, a promising accuracy is found.
Moving on to the \SI{100}{\giga\hertz} simulation data with channel estimates available every \SI{500}{\micro\meter}, as provided in \cref{fig:steps_ideal_100GHz}.
Here, we observe a deviation of up to \SI{14}{\micro\meter} in the worst case, whereas it is typically below \SI{3.8}{\micro\meter}.
The error, however, seems more like a constant offset for this y-axis resolution.
One can conclude from this that the sensing at the lower frequency might be less suitable, e.g., due to a different constellation of multipath components impacting the measurements.
Nonetheless, the error is orders of magnitude smaller than the motion, thus still highlighting that frequencies above \SI{100}{\giga\hertz} are suitable for the proposed monostatic monitoring of patients.

\subsection{Sensitivity Analysis: Impact of Patient Pose}
\label{ch:eval_B}

\begin{figure}[t!]
	\setlength\fwidth{0.9325\linewidth}
	\setlength\fheight{0.45\fwidth}
	\centering
% Thesis colors:
\definecolor{mycolor1}{rgb}{0.894118,0.101961,0.109804}% red
\definecolor{mycolor2}{rgb}{0.344117,0,0.69803}% indigo
\definecolor{mycolor3}{rgb}{0.71, 0.4, 0.1}% brown
\definecolor{mycolor4}{rgb}{0.00000,0.4400,0.85100}% blue
\definecolor{mycolor5}{rgb}{0.301961,0.686275,0.290196}% green
\definecolor{mycolor6}{rgb}{0.25, 0.25, 0.28}% gray/black

\begin{tikzpicture}

\begin{axis}[%
width=0.951\fwidth,
height=\fheight,
at={(0\fwidth,0\fheight)},
scale only axis,
xmin=-5,
xmax=185,
xlabel style={font={\footnotesize \color{white!15!black}}, yshift=0.75mm},
xlabel={Channel Estimate Index},
ymin=-2,
ymax=31.25,
ylabel style={font={\footnotesize \color{white!15!black}}},
ylabel={Tracked Motion $\left[\si{\milli\meter}\right]$},
axis background/.style={fill=white},
xmajorgrids,
ymajorgrids,
legend style={at={(0.026,0.9625)}, anchor=north west, legend cell align=left, align=left, draw=white!15!black,font=\footnotesize, row sep=-0.02cm, inner xsep=2pt, inner ysep=0pt},
legend columns=1,
minor x tick num=3,
minor y tick num=4,
ticklabel style={font=\scriptsize},
]
%% Annotation
% upper ellipse
\draw[color=mycolor6, line width = 1.25pt, dash dot, rotate around={-64.5:(95,15.9)}] (95,15.9) ellipse (3 and 7.5); 
\node[anchor=south, xshift=2.25mm, yshift=8mm] at (95,15.9) {
	\scriptsize
	\textcolor{mycolor1}{
		\shortstack{
			\contour{white}{Accurate Monitoring	of Gradual}\\
			\contour{white}{Movement at Ideal Conditions}
		}
	}
};
\draw[->, line width=1pt, color=mycolor6] (95,22.9)--(99.8,17.5);
% lower ellipse
\draw[color=mycolor6, line width = 1.25pt, dash dot, rotate around={-90:(150,0)}] (150,0) ellipse (3 and 7.5); 
\node[anchor=south, xshift=-12.5mm, yshift=0mm] at (150,0) {
	\scriptsize
	\contour{white}{
		\textcolor{mycolor5}{
			Fully Orthogonal Motion Untraceable
		}
	}
};

%% LINES
\addplot [color=black, line width=1.0pt]
  table[row sep=crcr]{%
1	0.166666666666657\\
180	29.9999999999999\\
};
\addlegendentry{Expectation}

\addplot [color=mycolor1, densely dashed, line width=0.75pt]
table[row sep=crcr]{%
	1	0.166189122976789\\
	155	25.8243738276068\\
	172	28.6560611979703\\
	180	29.9892305627157\\
};
\addlegendentry{No Rotation}

\addplot [color=mycolor2, densely dashed, line width=0.75pt]
table[row sep=crcr]{%
	1	0.0602848061421923\\
	6	0.362128029315954\\
	7	0.331866351036467\\
	14	0.76061871030646\\
	15	0.915812714775996\\
	20	1.21850981114719\\
	21	1.38591832108517\\
	28	1.99203942327142\\
	29	1.76459984323765\\
	49	2.97777073693362\\
	50	2.9054833122434\\
	71	4.1956380172596\\
	72	4.3739196880546\\
	86	5.21711263664369\\
	109	6.6106650808982\\
	127	7.6974897022109\\
	128	7.73518496521291\\
	161	9.73207041978014\\
	179	10.820358402467\\
	180	10.5718276056775\\
};
\addlegendentry{$\gamma=\SI{22.5}{\degree}$}

\addplot [color=mycolor3, densely dashed, line width=0.75pt]
table[row sep=crcr]{%
	1	0.110880987152655\\
	33	3.64511164493894\\
	34	3.55388680698854\\
	51	3.55388680698854\\
	52	3.66273964545445\\
	53	3.55388680698854\\
	85	3.55388680698854\\
	86	3.48246432944975\\
	92	3.48611677855561\\
	98	3.48238708152292\\
	102	3.48237398586431\\
	103	3.55388680698854\\
	124	3.55388680698854\\
	125	3.45424833889808\\
	129	3.93490782126193\\
	130	4.10461100862349\\
	131	4.10439550078033\\
	132	4.0535409036552\\
	180	4.0535409036552\\
};
\addlegendentry{$\gamma=\SI{45.0}{\degree}$}

\addplot [color=mycolor4, densely dashed, line width=0.75pt]
table[row sep=crcr]{%
	1	-0\\
	17	-0\\
	18	0.247841478027169\\
	41	2.51925470575873\\
	42	2.34031810652112\\
	92	7.77160291543723\\
	93	7.66890467713634\\
	111	9.44718763594847\\
	112	9.45948393727653\\
	159	14.5641638843456\\
	160	14.4899688033333\\
	164	14.4899688033333\\
	165	14.2610751793336\\
	180	15.7430904514598\\
};
\addlegendentry{$\gamma=\SI{60.0}{\degree}$}

\addplot [color=mycolor5, densely dashed, line width=0.75pt]
table[row sep=crcr]{%
	1	-0.000396832188471308\\
	3	0.0315867223206112\\
	5	0.0162360962276296\\
	7	-0.00218776476191351\\
	10	0.0312070391782129\\
	13	-0.00184290921757224\\
	14	0.00346161935846112\\
	16	0.0323056112545714\\
	20	-0.000534538919566785\\
	22	0.0314931731908814\\
	24	0.0164067131870809\\
	26	-0.0021962250468448\\
	29	0.0312439248049827\\
	33	0.0033634967862497\\
	35	0.0322987185469685\\
	39	-0.00052742456458077\\
	41	0.0314733349890446\\
	43	0.0164117033281173\\
	45	-0.00218239186480673\\
	48	0.0304534391966058\\
	100	0.0304534391966058\\
	101	-0.00639898774497283\\
	102	-0.00639898774497283\\
	103	0.0304534391966058\\
	180	0.0304534391966058\\
};
\addlegendentry{$\gamma=\SI{90.0}{\degree}$}

\coordinate (annotate_empty) at (0,0);
\coordinate (annotate_full) at (180,0);

\end{axis}

\node[align=left, font=\notsotiny, yshift=-7mm, xshift=3.85mm] at (annotate_empty) {(Lung Empty)};
\node[align=right, font=\notsotiny, yshift=-7mm, xshift=-2.6mm] at (annotate_full) {(Lung Full)};

\end{tikzpicture}%
	\vspace{-7.5mm}
	\caption{
		Measured patient motion for different poses at \SI{300}{\giga\hertz} based on continuous cumulation of measured motion steps, cf.~\cref{fig:steps_ideal_300GHz}.
	}
	\label{fig:trajs_all_300GHz}
\end{figure} 
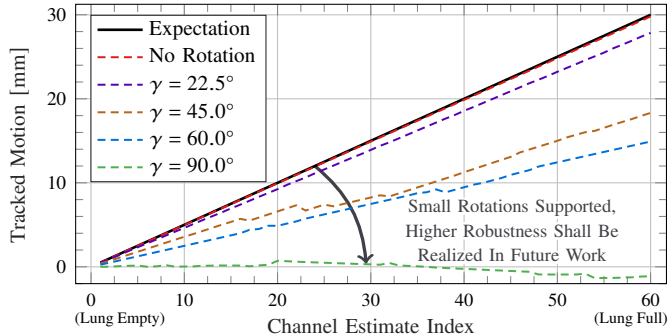
\begin{figure}[t!]
	\vspace{-2.5mm}
	\setlength\fwidth{0.9325\linewidth}
	\setlength\fheight{0.45\fwidth}
	\centering
% Thesis colors:
\definecolor{mycolor1}{rgb}{0.894118,0.101961,0.109804}% red
\definecolor{mycolor2}{rgb}{0.344117,0,0.69803}% indigo
\definecolor{mycolor3}{rgb}{0.71, 0.4, 0.1}% brown
\definecolor{mycolor4}{rgb}{0.00000,0.4400,0.85100}% blue
\definecolor{mycolor5}{rgb}{0.301961,0.686275,0.290196}% green
\definecolor{mycolor6}{rgb}{0.25, 0.25, 0.28}% gray/black
\begin{tikzpicture}

\begin{axis}[%
width=0.951\fwidth,
height=\fheight,
at={(0\fwidth,0\fheight)},
scale only axis,
xmin=-1.65,
xmax=61.7,
xlabel style={font={\footnotesize \color{white!15!black}}, yshift=0.75mm},
xlabel={Channel Estimate Index},
ymin=-2,
ymax=31.25,
ylabel style={font={\footnotesize \color{white!15!black}}},
ylabel={Tracked Motion $\left[\si{\milli\meter}\right]$},
axis background/.style={fill=white},
xmajorgrids,
ymajorgrids,
legend style={at={(0.025,0.9625)}, anchor=north west, legend cell align=left, align=left, draw=white!15!black,font=\footnotesize, row sep=-0.02cm, inner xsep=2pt, inner ysep=0pt},
legend columns=1,
minor x tick num=3,
minor y tick num=4,
ticklabel style={font=\scriptsize},
xtick={0, 10, 20, 30, 40, 50, 60},
xticklabels={{0}, {10}, {20}, {30}, {40}, {50}, {60}},
]
%% Annotation
\draw[->, line width=1.25pt, color=mycolor6] (24,12) to[bend right=-22.5] (29.5, 0.375);
\node[anchor=south] at (45.5,-0.5) {
	\scriptsize
	\textcolor{mycolor6}{
		\shortstack{
			\contour{white}{Small Rotations Supported,}\\
			\contour{white}{Higher Robustness Shall Be}\\
			\contour{white}{Realized In Future Work}
		}
	}
};

%% LINES
\addplot [color=black, line width=1.0pt]
  table[row sep=crcr]{%
	1	0.5\\
	60	30\\
};
\addlegendentry{Expectation}

\addplot [color=mycolor1, densely dashed, line width=0.75pt]
table[row sep=crcr]{%
	1	0.496425303714958\\
	2	0.992845825853912\\
	3	1.50680058170412\\
	60	29.7964464435946\\
};
\addlegendentry{No Rotation}

\addplot [color=mycolor2, densely dashed, line width=0.75pt]
table[row sep=crcr]{%
	1	0.464758139437308\\
	3	1.38098378368757\\
	4	1.86782526621922\\
	8	3.70905784075984\\
	9	4.18097685621274\\
	10	4.63097072743499\\
	11	5.11428184902292\\
	12	5.57967877375581\\
	14	6.47336315997518\\
	15	6.95056298571905\\
	16	7.39958884055103\\
	18	8.33414762474659\\
	20	9.24659090672768\\
	21	9.75268762201389\\
	23	10.6346262849064\\
	24	11.1284208568979\\
	26	12.0536203708193\\
	28	12.9920913575499\\
	29	13.4340484066637\\
	31	14.4025416851847\\
	32	14.8221323351129\\
	38	17.6405627056565\\
	39	18.0733953754273\\
	40	18.5734440317633\\
	41	19.0186379448204\\
	43	19.9581438200846\\
	46	21.3725960507192\\
	47	21.8552895560027\\
	48	22.2798187486055\\
	49	22.7299177158972\\
	50	23.2157205118133\\
	52	24.1260450362595\\
	53	24.6040283763825\\
	55	25.4958130131811\\
	56	26.0150184764792\\
	57	26.4573174886944\\
	58	26.9186893650227\\
	59	27.4022654403324\\
	60	27.8454749105853\\
};
\addlegendentry{$\gamma=\SI{22.5}{\degree}$}

\addplot [color=mycolor3, densely dashed, line width=0.75pt]
table[row sep=crcr]{%
	1	0.374285824075429\\
	3	1.07350218989078\\
	10	3.58025968550891\\
	11	3.9183060465911\\
	14	5.00361969609981\\
	15	5.34490155359408\\
	16	5.70853071462803\\
	17	5.50402695802276\\
	22	7.32283835327856\\
	23	6.69854560866585\\
	25	7.39884508282017\\
	26	7.20776547317271\\
	31	8.51492498378121\\
	32	8.38360805314814\\
	39	10.8727000802234\\
	40	11.2635978837411\\
	45	13.0910729219916\\
	46	13.6199869423387\\
	50	14.9985351143389\\
	51	15.3986571543499\\
	53	16.0277082829958\\
	54	16.1787535826521\\
	56	16.9178448847521\\
	58	17.5962378221442\\
	59	17.9728369910972\\
	60	18.3200237685377\\
};
\addlegendentry{$\gamma=\SI{45.0}{\degree}$}

\addplot [color=mycolor4, densely dashed, line width=0.75pt]
table[row sep=crcr]{%
	1	0.250214865364249\\
	16	3.99609761540874\\
	17	4.40523543333273\\
	18	4.50489488347387\\
	19	4.88955622020392\\
	20	4.88438095428381\\
	22	5.4061135690722\\
	23	5.73841375524923\\
	34	8.4837680996402\\
	35	8.76594600861899\\
	36	8.8810401163098\\
	37	9.2185041119728\\
	38	8.86417828106389\\
	41	9.77669659762797\\
	42	9.98591947387108\\
	44	10.5715062518448\\
	45	10.9427123422645\\
	47	11.510231339808\\
	48	11.9414497866553\\
	60	14.9154366206944\\
};
\addlegendentry{$\gamma=\SI{60.0}{\degree}$}

\addplot [color=mycolor5, densely dashed, line width=0.75pt]
table[row sep=crcr]{%
	1	-0.000655731681170835\\
	2	0.0269332056457614\\
	4	0.125080303871101\\
	5	0.143354883506873\\
	6	0.0155892448010988\\
	7	0.0550127745113684\\
	8	0.203499958154438\\
	9	0.0204972356857027\\
	11	0.0555209687693008\\
	12	0.123635411377151\\
	14	0.159442318145167\\
	16	0.150119039103579\\
	17	0.0660372219594834\\
	18	0.169606991972209\\
	19	0.245563396444112\\
	20	0.715267068805154\\
	23	0.566617138172674\\
	28	0.36852003693906\\
	31	0.253058472054654\\
	32	0.456665789387458\\
	33	0.162295449307102\\
	35	0.0617000600790831\\
	36	0.00116559463709365\\
	37	-0.0897838705762695\\
	47	-0.592383300569388\\
	48	-0.910859720261691\\
	52	-0.91428635234567\\
	53	-0.820849491389573\\
	54	-1.34082257384696\\
	56	-1.31714707040656\\
	58	-1.25016702396054\\
	59	-1.15433826907424\\
	60	-1.12650887853161\\
};
\addlegendentry{$\gamma=\SI{90.0}{\degree}$}

\coordinate (annotate_empty) at (0,0);
\coordinate (annotate_full) at (60,0);

\end{axis}

\node[align=left, font=\notsotiny, yshift=-7mm, xshift=3.85mm] at (annotate_empty) {(Lung Empty)};
\node[align=right, font=\notsotiny, yshift=-7mm, xshift=-2.6mm] at (annotate_full) {(Lung Full)};

\end{tikzpicture}%
	\vspace{-7.5mm}
	\caption{
		Measured patient motion for different poses at \SI{100}{\giga\hertz}.
	}
	\label{fig:trajs_all_100GHz}
	\vspace{-3.75mm}	
\end{figure}

In the second part of the evaluation, we consider the impact of the user rotation angle on breathing motion monitoring.
For the \SI{300}{\giga\hertz} simulation data depicted in \cref{fig:trajs_all_300GHz}, we find that the overall trajectory measurement is significantly distorted if the user has a non-ideal pose.
In the worst case of $\gamma=\SI{90}{\degree}$ the behavior is as expected in \cref{ch:conceptualization_B}, i.e., no motion is observed.
Thus, the need for our proposed technique to reconstruct the lost motion information is confirmed.
The trajectories observed for $\gamma\in\left[\SI{22.5}{\degree},\SI{45.0}{\degree},\SI{60.0}{\degree}\right]$ are furthermore mixed up and do not match the behavior predicted by \cref{eq:3}.
This could be due to multipath distorting the channel estimates.
Using the \SI{100}{\giga\hertz} carrier, the motion trajectories in \cref{fig:trajs_all_100GHz} more closely resemble a linear function with the slope depending on $\gamma$.
Here we find that small pose changes of, e.g., up to \SI{22.5}{\degree} rotation do not significantly impact the breathing motion tracking.
As such, the system will not require an otherwise motion\-less patient, e.g., monitoring during sleep could be feasible.
Nonetheless, for generalization of the system in terms of, e.g., patient pose  (position and rotation) and room layout, a tracking enhancement using the concept derived in \cref{ch:conceptualization} is necessary.
\section{Outlook on Future Work}
\label{ch:concl}

In this study, we investigated the use of 6G networks for health monitoring purposes. 
A concept for breathing rate tracking in patient rooms was introduced using \ac{THz} channel phase information from monostatic sensing.
The conducted case study revealed that the motion of the patient's chest can be measured with an error of less than \SI{1}{\micro\meter} at \SI{300}{\giga\hertz}, but only if the motion is aligned with the \ac{LOS} path.
This work outlined in detail how multiple propagation paths, together with bearing information, can be used to reconstruct lost orthogonal motion components.
As such, our ongoing work aims to confirm this by facilitating low-cost \ac{RIS}-based \ac{VLOS} propagation paths.
A novel network planning scheme shall be developed afterward to mount the \acp{RIS} at optimized positions such that any patient pose can be accommodated.

\footnotesize %% No need to waste space for acknowledgement and references
%% ACKNOWLEDGEMENT SECTION
\phantomsection
\addcontentsline{toc}{section}{Acknowledgment}
\section*{Acknowledgment}
\vspace{-0.5mm}
{	
	\begin{spacing}{0.97875}
		This work has been partly funded by the German Federal Ministry of Education and Research (BMBF) in the course of the \emph{6GEM Research Hub} under the grant number 16KISK038.
		Further funding has been received from the German Research Foundation (DFG) Project-ID 287022738 \emph{TRR 196} for \emph{Projects M01} and \emph{S04}.
		Additional support has been given by the Ministry of Economic Affairs, Industry, Climate Action, and Energy of the State of North Rhine-Westphalia (MWIKE NRW) along with the \emph{Competence Center 5G.NRW} under the grant number 005-01903-0047.
		This work has~also received funding from the program \emph{Netzwerke 2021}, an initiative of the Ministry of Culture and Science of the State of North Rhine-Westphalia~(MKW~NRW).
	\end{spacing}
}

%% REFERENCES SECTION
% Correct erroneous PDF bookmark (spelling + hyperlink) - Part 2
\phantomsection
\addcontentsline{toc}{section}{References}
\bibliographystyle{IEEEtran}

\begin{spacing}{0.95466}
	\fontdimen2\font=0.5915ex % inter word space
	\sloppypar
	\bibliography{IEEEabrv,bibliography}
	\fontdimen2\font=\origiwspc
\end{spacing}

\end{document}